# Resource Utilization Monitoring for Raw Data Query Processing

Mayank Patel
*Distributed Databases Group*
*DA-IICT*
Gandhinagar, Gujarat, India
mayank@daiict.ac.in

Minal Bhise
*Distributed Databases Group*
*DA-IICT*
Gandhinagar, Gujarat, India
minal_bhise@daiict.ac.in

***Abstract*—Scientific experiments, simulations, and modern applications generate large amounts of data. Data is stored in raw format to avoid the high loading time of traditional database management systems. Researchers have proposed many techniques to improve query execution time for raw data and reduce data loading time for traditional systems. The core of all the proposed techniques is efficient utilization of resources by processing only required data or reducing operations on data. The processed data caching in the main memory or disk can resolve this issue and avoid repeated processing of data. However, limitations of resources like main memory space, storage IO speeds, and additional storage space requirements on disk need to be considered to provide reliable and scalable solutions for cloud or in-house deployments. This paper presents improvements to the raw data query processing framework by integrating a resource monitoring module. The experiments were performed using a scientific dataset known Sloan Digital Sky Survey (SDSS). Analysis of monitored resources revealed that sampling queries had the lowest resource utilization. The PostgresRAW can answer simple 0-JOIN queries faster than PostgreSQL. While one or more JOIN complex queries need to be answered using PostgreSQL to reduce workload execution time (WET). The results section discusses resource requirements of simple, complex, and sampling type queries. The result analysis of query types and resource utilization patterns assisted in proposing Query Complexity Aware (QCA) and Resource Utilization Aware (RUA) data partitioning techniques for raw engines and DBMS to reduce cost or data to result time.***

## I. INTRODUCTION

Scientific experiments are generating massive amounts of data every day. The Large Hadron Collider LHC experiment at CERN generates 90PB of data every year [1]. The amount of data processed by CERN servers reach 1PB per day. The astronomy observations datasets like Sloan Digital Sky Survey SDSS have generated data reaching 273TB [2]. Weather satellites of NASA from the Earth Observing System (EOS) project generate more than 3.3 TB of data every day [3]. The traditional way of query processing raw data required the entire dataset to be loaded into DBMS. The data loading time DLT is very high for most DBMS as these systems parse, tokenize and convert the data to a specific database format. The use of pre-processed data allows DBMSs to answer queries faster.

The dataset size of scientific and modern applications forces distributed computing approaches in cloud environments or private data servers. The computer parts at these data centers are prone to failures. Therefore, replication and other fail-safe measures are taken to keep the applications running around the clock. Consequently, the current deployment system involves more resources and incurs more costs. Researchers have observed that only 12% of CPUs were utilized at data centers [4]. The underutilization was due to the fact that most database management systems DBMSs were utilizing only 25-50% CPU resources. Fig. 1 displays CPU utilization by different DBMS like Shore-MT, DBMS D, VoltDB, and HyPer [4]. Therefore, monitoring CPU utilization is important to identify underutilization and release additional resources to reduce application costs.

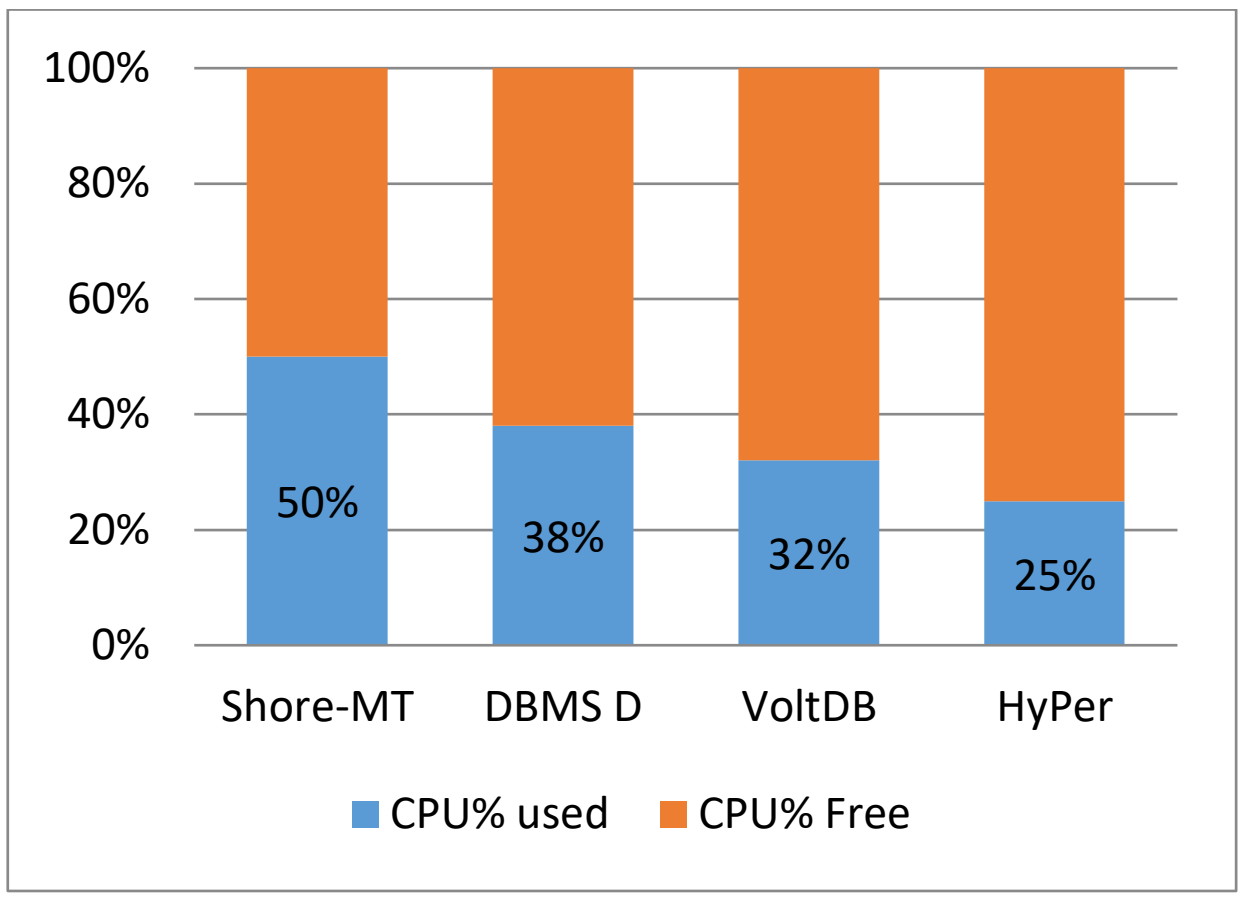

Fig. 1. CPU Utilization in DBMSs [4]

The raw or in-situ engines query raw data without loading it into DBMS. The query execution time is high in raw engines due to the data processing task transpiring after the arrival of the query. Many engines keep the processed data into memory and build indices to help future queries [5], [6]. However, the cached data gets removed as future queries need new data. Such techniques require knowledge of main memory utilization due to cached data and run-time work memory requirements of a query to improve the performance of the proposed technique. Online analytical processing OLAP queries require processing of historical data, which had raised the reparsing issue for raw data query processing. Researchers have proposed incremental data loading techniques into DBMS to avoid upfront utilization of resources & infrastructure requirement costs [7], [8]. The query processing is read-intensive, while data loading tasks are write-intensive. Researchers had observed that disk IO is available when query data is cached in the main memory. This allowed researchers to propose incremental data loading and indexing techniques. These techniques have been discussed in section II.

The CPU, RAM, and IO monitoring have allowed researchers to propose data partitioning, task scheduling, and resource allocation techniques to reduce workload execution time (WET) and resource utilization. T. Chauhan et al. have also proposed to monitor resource utilization to match cloud service level agreements [9]. This paper tries to find the resource monitoring tools to set up a real-time resource monitoring framework (RMF) and incorporate it with the raw data query processing framework to analyze resource requirements of data processing tasks.

### A. Motivation

Researchers have observed resource utilization of CPU, RAM, and IO resources to propose efficient data processing techniques. Analyzing resource utilization patterns can allow us to find a resource-efficient way of processing the raw data.

### B. Problem statement:

Monitoring resource utilization can help us predict the amount of data a machine with a data processing tool can process without performing scaling experiments. Monitoring the actual resource requirements of a query may help us decide which tool is best suited for which type of query. However, it is essential to know how to monitor resource utilization. The details of available resource monitoring tools and the overhead of using those tools must not impose an additional burden on resources. This paper aims to develop a RMF to analyze resource utilization patterns for raw data query processing.

### C. Paper Contributions

- The paper proposed a resource monitoring framework (RMF).
- Integration of resource monitoring module with the raw data query processing framework.
- Used proposed framework for monitoring resource utilization for scientific data query processing.
- Monitored & compared resources used by the in-situ engine and state-of-the-art DBMS while processing a real-world scientific dataset.
- Discovered the amount of data a machine can process efficiently based on resource requirements of a tool.
- Recommend which tool is best for what type of workload queries after analyzing the resource requirements of different queries.
- Proposed two raw data partitioning methods. 1) Query Complexity Aware (QCA), and 2) Resource Utilization Aware (RUA).

## II. Related Work

This section discusses the research work that utilized resource monitoring to propose efficient raw data query processing techniques.

The data loading and query processing resource requirements differ significantly. The row stores, column stores, and graph based DBMSs process and organize the data differently, which directly impacts DLT & query execution time QET [10], [11], [12]. An experimental analysis of resources required by data loading tasks for traditional DBMSs has been discussed in a paper [13]. Researchers have observed resource utilization to find reasons for high data loading time and ways to reduce it. The parallel loading of data did not improve DLT time for disk-based storage. The paper determined that the storage medium is the bottleneck, and parallel access to the disk further increases DLT time. The query execution time QET also gets affected due to data loading operations accessing disk in parallel [14]. It was observed that permanent storage devices like hard disks perform best with sequential access [13]. To avoid parallel access to the disk, researchers have proposed monitoring the CPU and IO utilization to find the idle time [14], [15]. The techniques utilized the idle disk time to load data into DBMS. Researchers have proposed to increase RAM utilization after identifying that default protocols had a high false abortion rate for processes [16].

Researchers had observed that a fraction of a database could answer most workload queries [17]–[19]. Researchers have proposed horizontal, vertical, and hybrid partitioning techniques to reduce the query execution time [20], [21]. PDC-Query work applied data blocking and summarization techniques to process scientific datasets using object data management systems (ODMS) [19]. PDC-Query approach utilized 50% of main memory to cache partition summaries. Materialized views can reduce QET by 26-79% for complex multi-join RDF queries [22]. Partial loading technique proposed to vertically partition the raw data, keeping one partition into DBMS and second in a raw format based on access costs & limited storage resources [8]. A most recent work improved utilization of storage resources by considering actual storage requirements of hot partitions to be loaded in DBMS [18].

Main memory caching and indexing techniques need to monitor the available RAM resource to efficiently manage the cached data [6], [23]. Many raw data query processing techniques have proposed incremental loading and indexing of raw data into traditional DBMS systems to reduce upfront utilization of resources [7], [24]–[26]. Researchers have proposed loading of data processed by queries before deleting them from main memory to eliminate reparsing issues [7]. The cloud deployment or in-house deployment of applications can considerably reduce the application running cost by efficiently utilizing the available resources [27]. Additionally, vendor lock-in issue persists for different cloud service providers. Therefore, S. Kotecha et al. proposed to translate SQL queries to GQL to reduce application migration costs [28]. Researchers have also proposed utilizing the client resources to load selective data into DBMS [29]. It is crucial to identify resources requirements of different tools and queries to perform query processing tasks over raw files and databases to reduce application working costs [30]. Most techniques discussed here used analysis of monitored resources to propose the techniques which tried to reduce resources and data to result time. However, the number of research papers is limited that displays the resource requirements of different types of queries on different tools.

## III. Resource Monitoring

This section explains the role of resources in raw data processing, followed by the updated framework.

### A. Role of Resources

Processing raw data needs all three core resources, CPU, RAM & Storage devices. Processing raw data using in-situ or raw engines requires storage resources to read stored data to answer queries. CPU resource performs data parsing, conversion, and tokenizing tasks. The main memory caches the processed data required by the CPU for faster access. The traditional DBMS stores the processed data back to disk for future queries. When a query arrives, a query plan is made, which might require fetching data from the disk. If the required data is already cached in RAM, it can be processed faster. The time of storing or reading data from the permanent storage is directly affected by the IO speed of the storage hardware resource. The RAM resource provides the cached data to the CPU for processing. The time required by CPU in processing query operations and generating results depends on the CPU processing speed. If the required data is not in RAM, the CPU has to wait for data to be fetched from the disk, known as a cache miss. The percentage of time wasted in the wait of IO completion represents idle CPU time when work was pending. Slow storage speed and smaller RAM size can increase cache misses and CPU Wait. The high wait percentage indicates delays in data processing by the CPU and results in high data to result time. The RMF allows us to understand the resource utilization patterns and propose an efficient way of processing the data.

## B. *Resource Monitoring Framework (RMF)*

The Resource Monitoring Framework (RMF) consists of CPU, RAM & IO resource monitoring modules that access resource utilization data from resource monitoring (RM) tools like *system monitor, task manager*, *SAR*, *top*, *htop*, *iotop* & others. RM tools can display resource utilization details in GUI or text based interfaces. The RMF should be capable of sending CPU, RAM, and IO resource monitoring commands to an RM tool and collecting output for further analysis. Monitoring utilization of resources during data processing tasks is the primary goal of this work. Therefore, the data processing tasks need to be associated with RM output. Therefore, the resource monitoring module needs to be integrated with an earlier proposed raw data query processing framework to achieve the correlation [10].

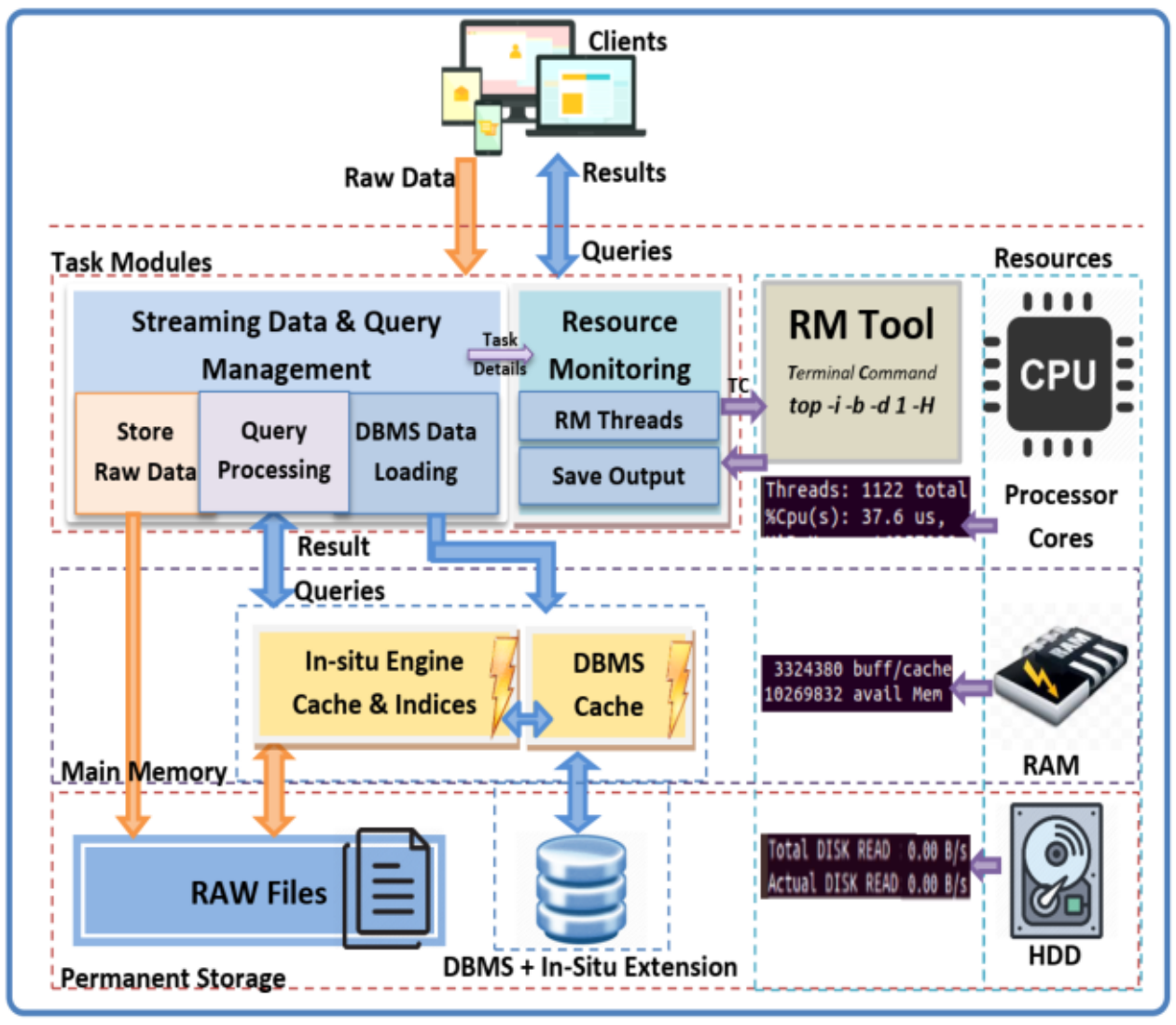

Fig. 2. Resource Monitoring Framework (RMF)

The updated raw data query processing framework with resource monitoring module can be seen in Fig.2. It can be observed that the RM modules send commands to the tools and receive the output. The received output is filtered to save monitored resources of the raw engine, DBMS, and framework processes. The module also saves total resource utilization but filters out details of other processes. At the same time, the RM modules get task details in real-time from the streaming data & query processing module. The task details and monitoring results are then combined in a list of comma-separated strings format and saved to the output result file. The output is not loaded to a database to avoid utilization of additional resources, which is a common practice used by developers. However, the real-time decision making algorithms require real-time analysis of recorded resource utilization. The following section discusses implementation of the RMF and incorporation of RM modules into the raw data query processing framework [10].

## C. *Algorithms & Data structures*

This section discusses experiment flow for resource monitoring experiments followed by pseudo-code.

### 1) *Data Structures*

The resource monitoring algorithm uses simple data structures like arrays and lists. The input workload file contents associate the task ID with task statements. Table I shows an example of a workload file containing task ID and task statements. The workload file is read and kept in a List<String[]> and read one by one by the query processing functions. The output data is stored in a List<String[]> before writing to a CSV result file. Table II shows a sample output list. The list is long as it stores all the CPU, RAM, and IO utilization results in total and filtered processes. It can be seen in Table II that the output list contains time to complete a task, CPU, RAM, and IO utilization during a TRUN task. The Postgres process CPU utilization is a single CPU core utilization while memory utilization is in % out of 100%.

TABLE I. WORKLOAD LIST

| **0[T_ID]** | **1[Statement]** |
|---|---|
| TRUN | "TRUNCATE TABLE PhotoPrimary;" |
| COPY | "COPY PhotoPrimary FROM '/…/PhotoPrimary.csv' (DELIMITER ',');" |
| Q0 | "Select count(objid) from PhotoPrimary;" |
| Q1 | "SELECT objID, ra ,dec FROM PhotoPrimary WHERE ra > 185 and ra < 185.1 AND dec > 56.2 and dec < 56.3 limit 100;" |

TABLE II. RESULT LIST

| **0** | **1** | **2** | **3** | **4** | **5** | **6** | **7...** |
|---|---|---|---|---|---|---|---|
| Q_ID | TRUN | Time | 444 | ms | #Rec. | 0 | star |
| CR_T | TRUN | Used | 3.9 | Free | 92.1 | IO | 1.9 |
| CR_P | Postgres | 1core | 79.4 | Mem | 0.2 | | |
| IO_T | TRUN | Read | 0.59 | Write | 0.02 | | |
| IO_P | Postgres | Read | 0.50 | Write | 0.01 | IO | 0.2 |

### 2) *Resource Monitoring Algorithms*

The raw data query processing algorithm is responsible for performing data loading and query processing tasks. This process sets the input and output file paths. It starts resource monitoring threads. The process sets and updates the currently running data processing task details in shared variables. The RM threads read those task details and incorporate them with RM output. RM thread tasks include execution of RM command on tool interface provided by OS terminal or RM tools. The resource utilization output stream is read by these threads using buffer reader. The RM threads filter the received resource monitoring output line by line to find specific resource utilization details and save it with task information in a CSV file. The filter parameters are set to find lines that contain *Postgres* and *Java* processes and their resource utilization. The *Postgres* process represents PostgresRAW and PostgreSQL processes, while *Java* processes are the framework tasks.

The RM output is not immediately stored to disk to avoid continuous usage of IO. The threads collect the RM information in the result list residing in main memory before flushing to the disk. The list collects the data until a predefined threshold value is met, i.e., the number of records filtered. The final result file contains the overall CPU, RAM & IO utilization observed during raw data processing. The resulting file also contains *Postgres* and *Java* process resource utilization information. The raw data query processing process interrupts both RM threads when the data processing tasks are completed. A few seconds of delay can be imposed to store collected data to disk. Raw data query processing and resource monitoring algorithm pseudo-codes have been presented below.

**Algorithm 1:** Raw Data Query Processing

**Data:** *w_p* = workload file path
*r_p* = result file path
*IsRM* = Monitoring threads flag
*RMType = CPU & RAM* or *IO* RM types
*RMcommand* = Monitoring command for tool
*RM_F* = Monitoring Freq.
*Task_ID* = data loading or Query ID

**Result:** Data Processing Task Execution Time & Results

**1.** ***RawDataQP*** *( w_p, IsRM, RM_F, r_p ):*
**2.** Set parameter values *w_p, r_p*
**3.** Set *RMcommand = top –I –b -d {RM_F} -H*
**4.** **If** *( IsRM == True )* **then**
    #Initialize. RM Threads A
**5.**     *RM_Thread ( RMType, RMcommand, r_p )*
**6.**     *RM_Thread.start ( )*
**7.**     *DB.Connect ( )*
    #start workload execution
**8.**     **for** each task *T* in *w_p* **do**
**9.**         Set static *Task_ID = T.T_ID*
**10.**         Results = *T.Statement.Execute ( )*
**11.**         Save time & Result count for each task
**12.**     **end**
**13.**     *RM_Thread.interrupt ( )*
**14.** Return;

**Algorithm 2:** Resource Monitoring

**Data:** *stdoutReader* = stores tool output

**Result:** Resource monitoring data with task ID

**1.** ***RM_Thread*** *( RMType, RMcommand, r_p ):*
**2.** Set local parameter values *T_ID*
    #Execute RM tool command
**3.** *Runtime.getRuntime ( ).exec( RMcommand )*
**4.** *BufferedReader stdoutReader = new BufferedReader (InputStreamReader ( getInputStream ( )))*
**5.** **While** *stdoutReader.readline* is NOT NULL
**6.**     **for** each *line* in *stdoutReader* **do**
**7.**         Filter RM data from *line*
**8.**         *RM_Info* = Set *T_ID* to RM monitoring*;*
**9.**         *WriteRM ( RM_Info, r_p );* #save data
**10.**     **end**
**11.** Run till interrupt from *RawDataQP*
**12.** Exit.

## IV. Framework Setup

Most traditional DBMS do not consider real-time resource utilization due to statistics collection and analysis overhead. These databases are configured to work for general workload and datasets. The specialized tuning of resource allocation is done by database admins based on their knowledge of the application domain, workload requirements, and DBMS software for specific applications. This section discusses how resource monitoring tools work and their setup with minimal overhead.

### A. Resource Monitoring tools

The operating system interfaces with the CPU, RAM & IO hardware to process, retrieve, and store data. The system monitor or task manager is an inbuilt resource monitoring application that provides the user an interface to view resource utilization. These inbuilt system monitoring tools may not have a customizable output based on our requirements. Therefore, installation of additional tools is required. We had checked *top*, *iotop*, *SAR*, *Sysstat*, *iostat*, *iosnoop*, and a few other resource monitoring tools that provide customizable resource utilization outputs [13]. We tried to find a single tool that can provide all required information and combine top & iotop output in a single command. But we could not find any such tool with the required details nor a combined command worked during framework implementation using java.

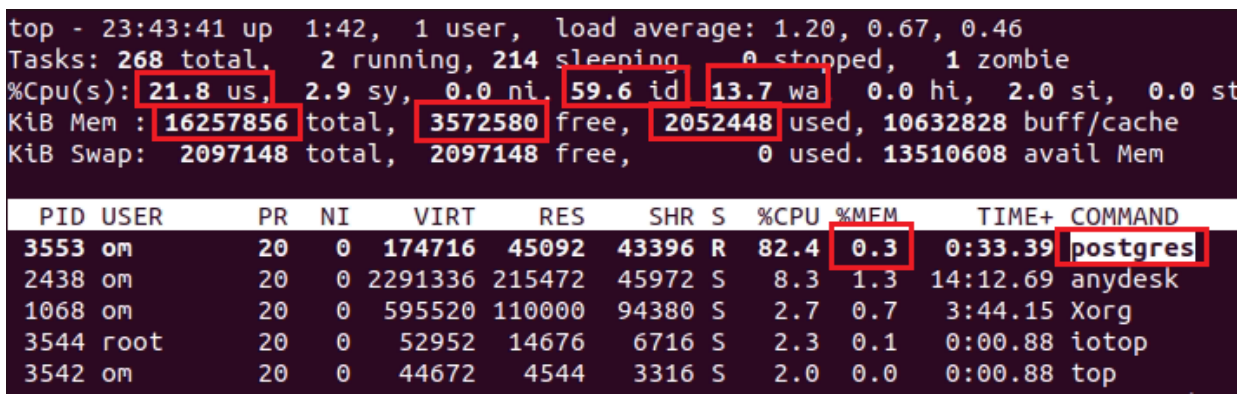

a) *top* tool: CPU & RAM

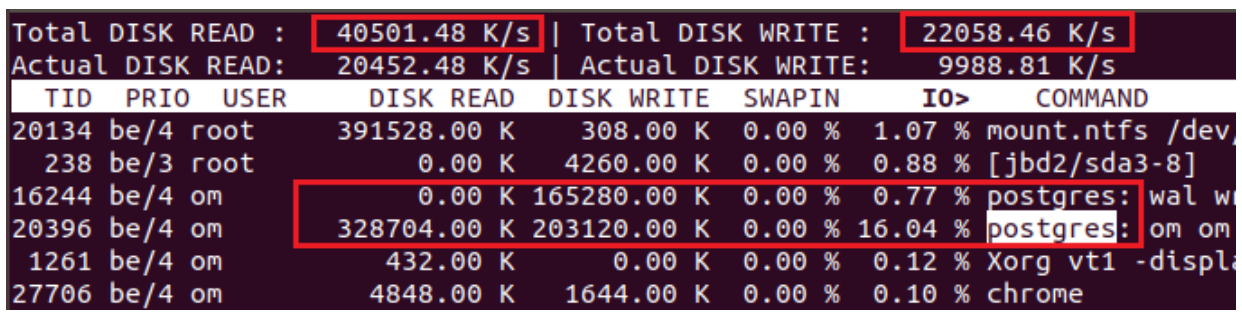

b) *iotop* tool: Disk IO

Fig. 3. Resource Monitoring Tool Output

For experiment purposes, *top* and *iotop* tools are used for their detailed output that matches our needs. The *top* tool provides total utilization and individual process utilization of CPU, RAM, and SWAP. The *iotop* provided IO utilization in % with total and individual process read/write in K/s. A sample resource monitoring output for both tools is shown in Fig 3. Fig 3 a) shows the *top* tool output with CPU, RAM, and important processes marked with red rectangles. Fig 3 b) displays IO utilization by processes and total read/writes.

### B. *Tool setup*

The in-situ extension and DBMS do not provide resource utilization details for each data processing task. Therefore, a data passing interface between resource monitoring tools and framework is required. The RM modules link data processing tasks with resource utilization outputs. The modules use *top* and *iotop* resource monitoring tools to monitor real-time resource utilization. These tools are installed on the operating system Ubuntu. RM threads pass terminal commands TC to the external RM tools and receive output. The output stream contains a lot of information. Storing all the information can quickly increase the result size. The basic experiments created huge result output files that reached a few GB in size containing only 3-6 hours of resource monitoring results. Therefore, the RM modules filter the required CPU, RAM, and IO utilization values for specific processes before saving. This output is then combined with current data processing task information, i.e., query id, before saving to the result file. The task ID with resource utilization helps correlate the resources required by each data processing task.

## V. Experimental Setup

This section describes the experimental setup consisting of Hardware & Software Setup, Dataset & Query Set.

### A. *Hardware & Software Setup*

The experiment machine included a quad-core Intel i5-6500 CPU. The max CPU speed is 3.20GHz. It had 16GB of RAM & in-built graphics Intel HD Graphics 530. The operating system of the machine is a 64-bit Ubuntu 18.04 LTS. A SATA hard disk drive with 500GB of storage space is used to permanently store raw files and databases. The disk rotation speed is 7200 rotations per minute. Ubuntu terminal was used to install *top* and *iotop* resource monitoring tools. The terminal provided the interface between the RMF and OS to monitor hardware resources. The RM tool commands were executed on the terminal using java code. The raw data query processing framework is built using java on eclipse[10]. The framework used PostgreSQL as a database management system to load raw data and PostgresRAW extension as the in-situ engine to execute queries directly on raw files.

### B. *Dataset & Query Set*

The Sloan Digital Sky Survey SDSS dataset is a real-world astronomical dataset of stars, galaxies, and other sky objects. The recent data release DR-16 increased SDSS dataset size to 273TB [2]. SDSS DR-16 has 134 tables and 59 views. Analysis of the query workload of 0.4M showed that 55% of queries of the workload belong to the *PhotoPrimary* view. Therefore, we extracted 18.8GB of data from *PhotoPrimary* view for the experiment. The number of records in the extracted table is 4M. The SDSS keeps track of queries executed on the dataset and logs them in *SQLlogAll* table. We had extracted the top 1000 unique queries executed on *PhotoPrimary* view. These queries represented 51% workload of the DR-16. The queries have been grouped based on the similarity of the attributes & query type, forming 12 query groups. We had used one representative query from each group for experiments.

## VI. Results & Discussion

This section discusses the experiment results. For convenience, the PgSQL and PgRAW aliases have been used to represent PostgreSQL and PostgresRAW.

### A. *RM Tool Configuration*

The *top* and *iotop* tools have multiple configuration settings, including filtering output details, sorting, and delays before refreshing resource utilization. The delay *–d* setting controls resource monitoring frequency for both tools. The freq. for both tools needs to be equal to correlate the outputs for analysis. Fig 4. shows the graph of resource monitoring frequency and CPU utilization. It can be seen that with the increase of monitoring frequency, the CPU utilization increases for both tools. The CPU utilization does not increase more than 25% because the machine had four cores. The 25% utilization shows that the entire one core out of four is used to monitor resources from 1000 and above observations per second. The higher observation rate also increases result output file size. The *top* & *iotop* RM threads need to filter those monitored records before storing, which may utilize additional two CPU cores and increase IO utilization.

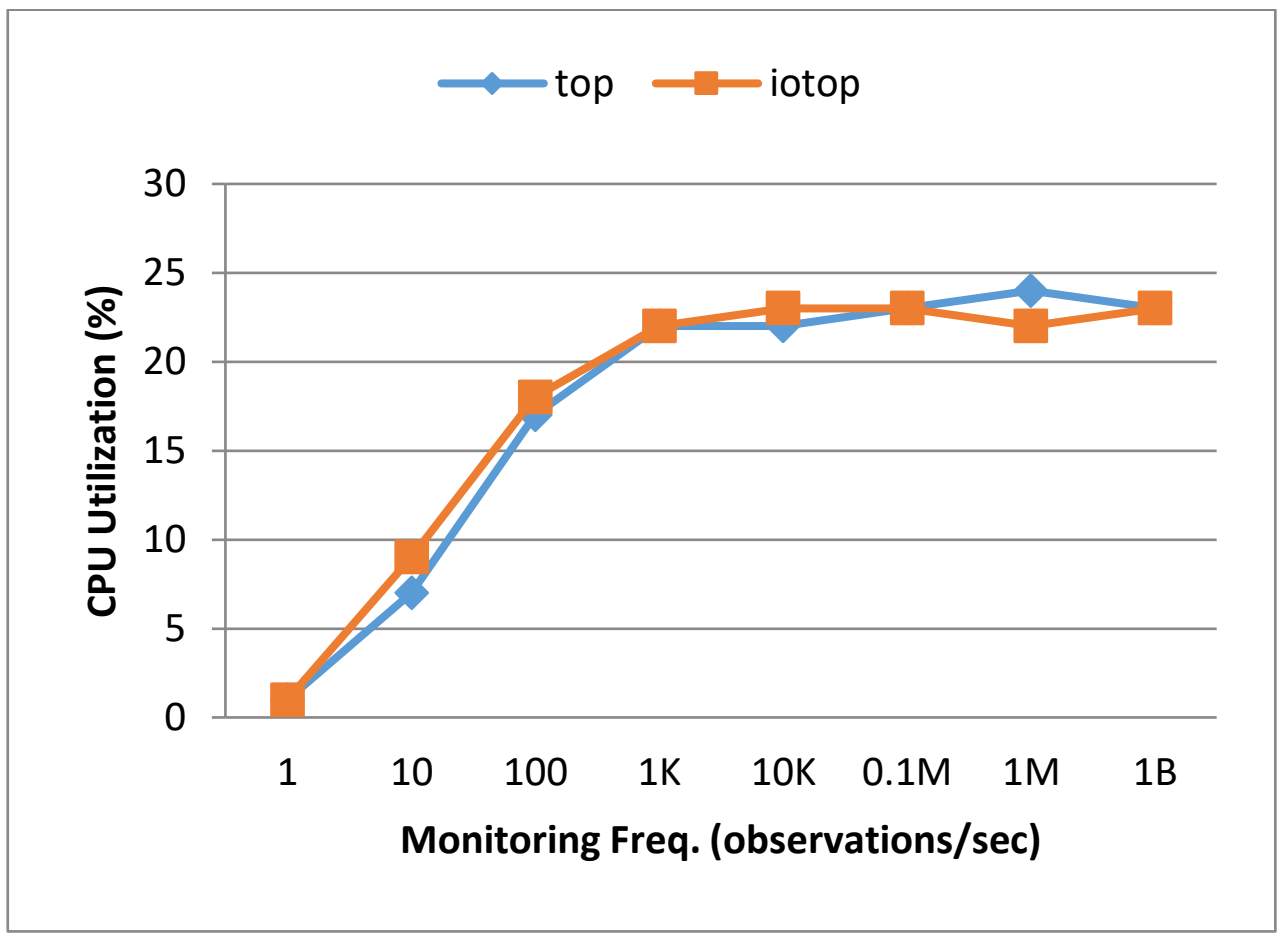

Fig. 4. RM Tools Overhead

CPU is the fastest resource in all resources. The CPU speed of 3.20GHz indicates that it can process $3.2x10^9$ instructions every second. The machine used for experiments is a 64bit, which means the CPU can process $3.2x10^9x64bit$ = 23.84GB of data every second. However, CPU is dependent on RAM to fetch data from the disk. Now, DDR3 RAM speed is less than a few GBs per second, while magnetic disks can only provide 300MB of data per second. While accessing smaller files, disks can provide only 30MB of data per second due to seek delay of 2-4ms for each file. This means the change in resource utilization of RAM & IO resources is slow, while CPU is dependent on them, which indicates the overall utilization change is not rapid. IO's 300MB per second data read speed represents less than 2% change in RAM resource utilization. Additionally, RM tools record resource utilization cumulatively, avoiding missed reading issues. Therefore, the monitoring frequency is set to 1 observation per second for most experiments to minimize resource monitoring overhead and keep resources available for actual workload processing.

### B. *Workload Execution on Raw*

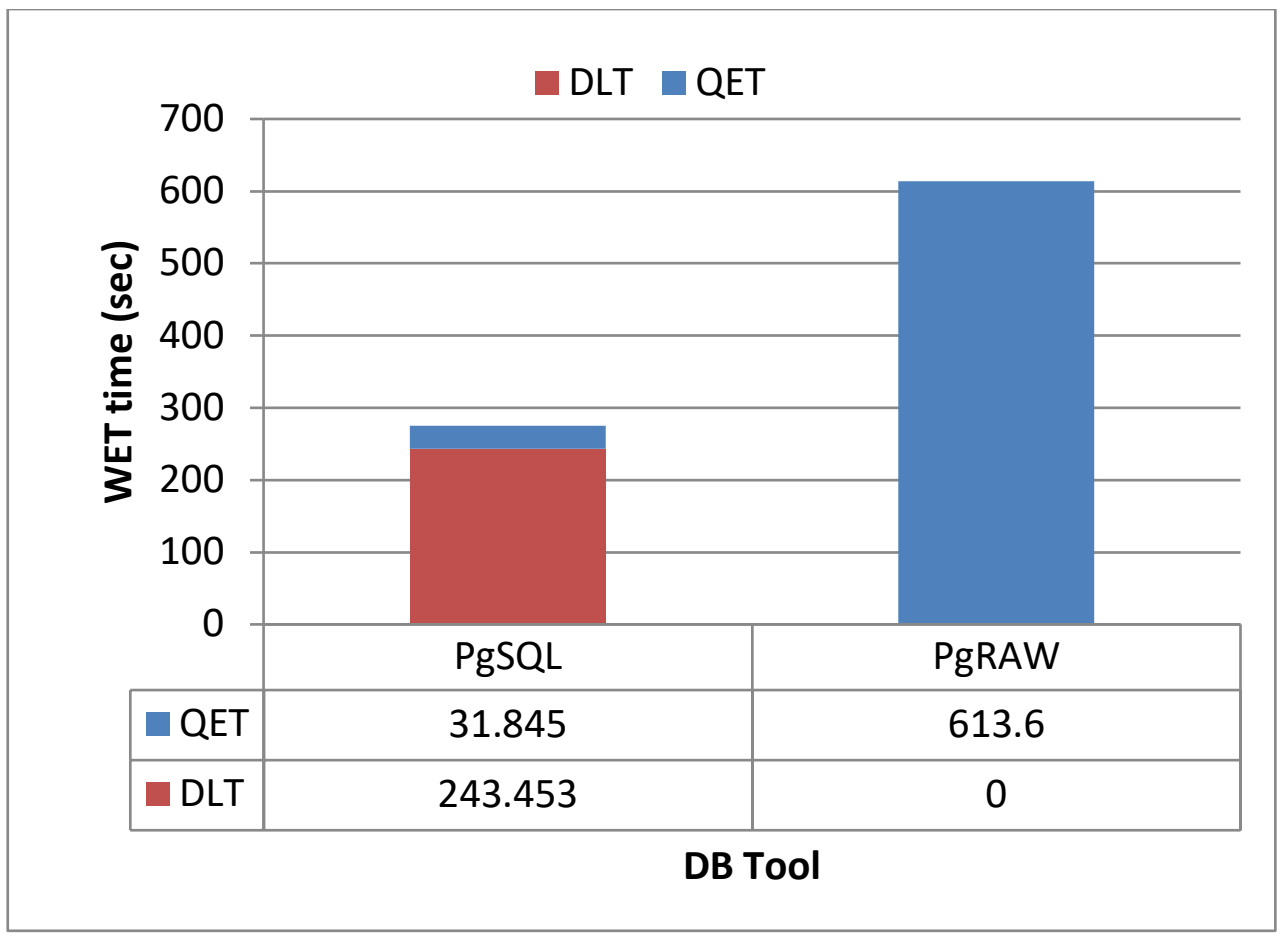

Fig. 5. WET for SDSS Dataset -1M

The raw data query processing framework uses PostgresRAW. It is a NoDB implementation to execute SQL queries directly on raw data [31]. It is an open-source tool. Fig 5 shows the comparison of WET time in PostgreSQL PgSQL with PostgresRAW PgRAW for the SDSS dataset having 1M records. The 1M records dataset required 4.6GB of disk space in raw format. It can be observed that raw data query processing engines like PgRAW have zero data loading time as they can start executing queries on the raw file immediately. At the same time, PgSQL required a considerable amount of time in loading data existing in a CSV file even with the fastest data loading technique COPY. However, PgSQL required only 5.19% time to execute 12 SDSS queries on loaded data compared to PgRAW. Therefore, it is crucial to load data into a database system to reduce QET of future queries.

### C. *QET for Query Types*

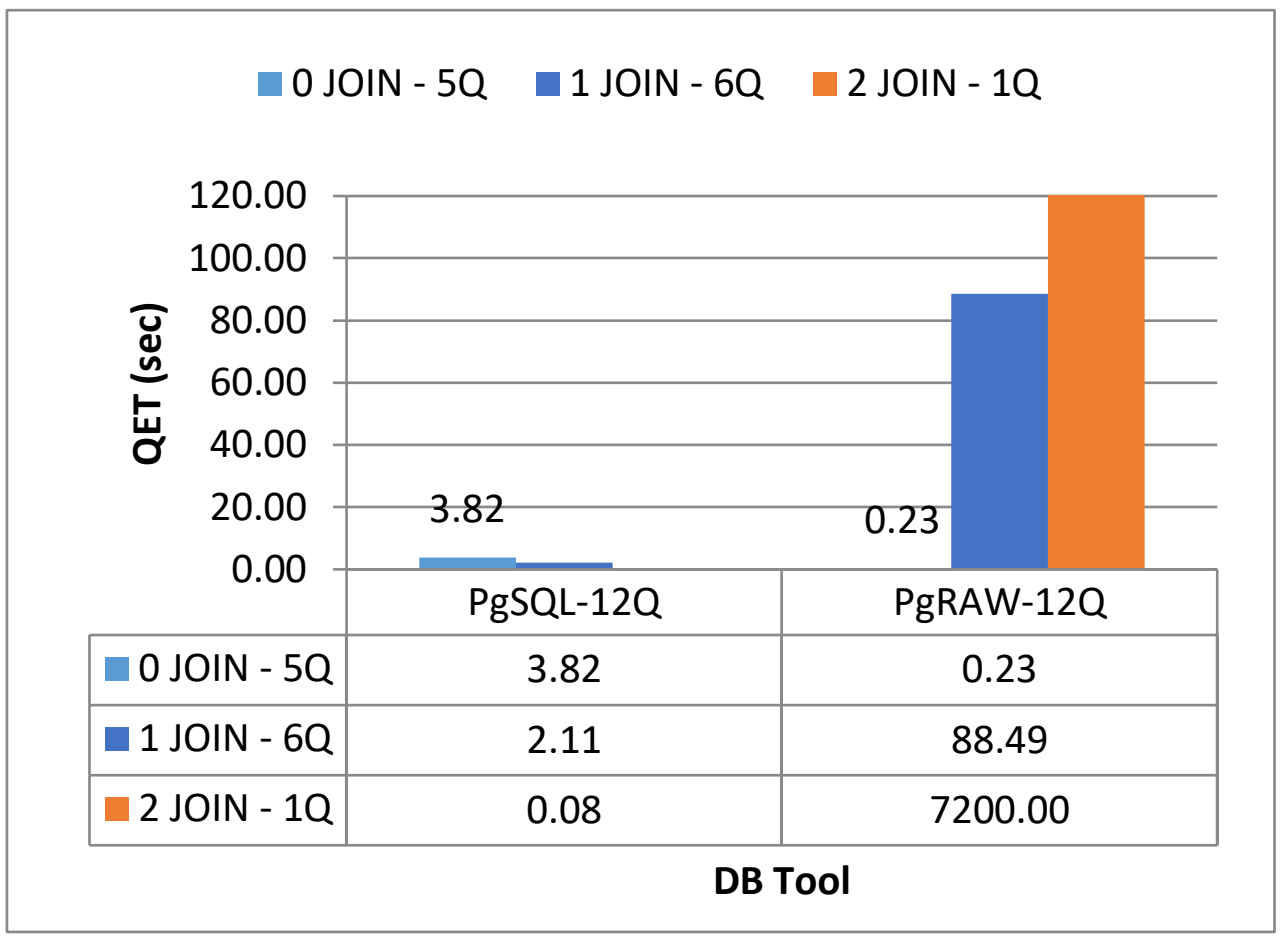

Fig. 6. SDSS Query Classification based on #JOIN

The comparison analysis of each SDSS workload query showed that some queries were faster in PgRAW. The detailed analysis of those SQL queries revealed that all of those queries were 0 JOIN simple queries. Fig. 6 compares the average query execution time of queries grouped by number of joins for PgSQL and PgRAW tools. It can be seen that the average QET of 0 JOIN queries in PgRAW is 93.9% lower as compared to PgSQL. However, 1 JOIN and 2 JOIN queries are much slower in PgRAW. For this experiment, all 1M dataset partition was used, which was pre-loaded and pre-cached before executing queries using PgSQL and PgRAW. A 2 JOIN query took more than 3 hours without any results in PgRAW, which shows that the tool is not optimized to handle 2 JOIN complex queries. The QET of this 2 JOIN complex query is not considered in most results to avoid highly skewed graphs.

From the query classification results shown in fig. 6, it can be determined that 0 JOIN simple queries can be executed using PgRAW while execution of 1 or more JOIN complex queries should be done using PgSQL after the dataset is loaded into a DBMS. Many researchers have proposed loading raw data into DBMS when IO resources are available. The resource monitoring experiments done in next section tries to find how much resources are utilized by both tools during workload execution. The analysis would provide insights into the possibilities of parallel execution of queries using PgRAW while data is loaded into PgSQL.

### D. *Resource Monitoring*

This section compares the resource utilization observed by RM tools for PgSQL and PgRAW.

#### 1) *Resource Utilization*

PgSQL and PgRAW follow different data processing techniques. This section provides insights into the actual resource requirements of these DB tools to process 1M records of SDSS dataset. PgSQL needs to load the entire dataset before query execution can start, while PgRAW starts the execution of queries immediately on raw data. Fig. 7 shows resource utilization graphs of both. The RM tools provided CPU utilization, IO utilization & CPU IO_WAIT in percentage, while the read and write bandwidth was calculated using total read and write observations considering max read speed as 300MB/sec and max write speed 200MB/sec.

The actual disk read and write observations were not used because read/write speed can be higher than actual data read/write. It would not have provided any valuable insights. The tools provided IO utilization for each process, so the plotted IO utilization is SUM of all *Postgres* and *Java* processes representing the raw data query processing framework. Fig. 7 a) shows resource utilization for PgSQL. It can be seen that CPU utilization during data loading tasks is ~20%. This is because data loading is an IO-dependent process. The CPU had to wait for IO for data to be read from the disk. The high IO utilization and high IO Wait spikes confirm that observation. It can be seen that once data loading task is completed, read and write to the disk are almost zero. It is because the entire dataset gets cached in the main memory to reduce QET of future queries. However, the IO and IO Wait readings reflect those changes a little later due to slower reflection of storage hardware utilization readings and *VACUUM* process reclaiming space delaying access to disk for most processes. The *VACUUM* process is the reason behind high write spike just after data loading is complete as it is removing old tuples, which facilitate faster disk access to the new data. The query processing tasks increased RAM utilization by 3.1% as the query processing tasks process the cached data in main memory to provide results. The query processing tasks utilized a single CPU core, which is 25% for a quad-core CPU. The *Postgres* processes like *wal*, *checkpointer process*, *parse*, *bind*, and other processes, including *Java* processes, could utilize up to 77% CPU resources but only for one second.

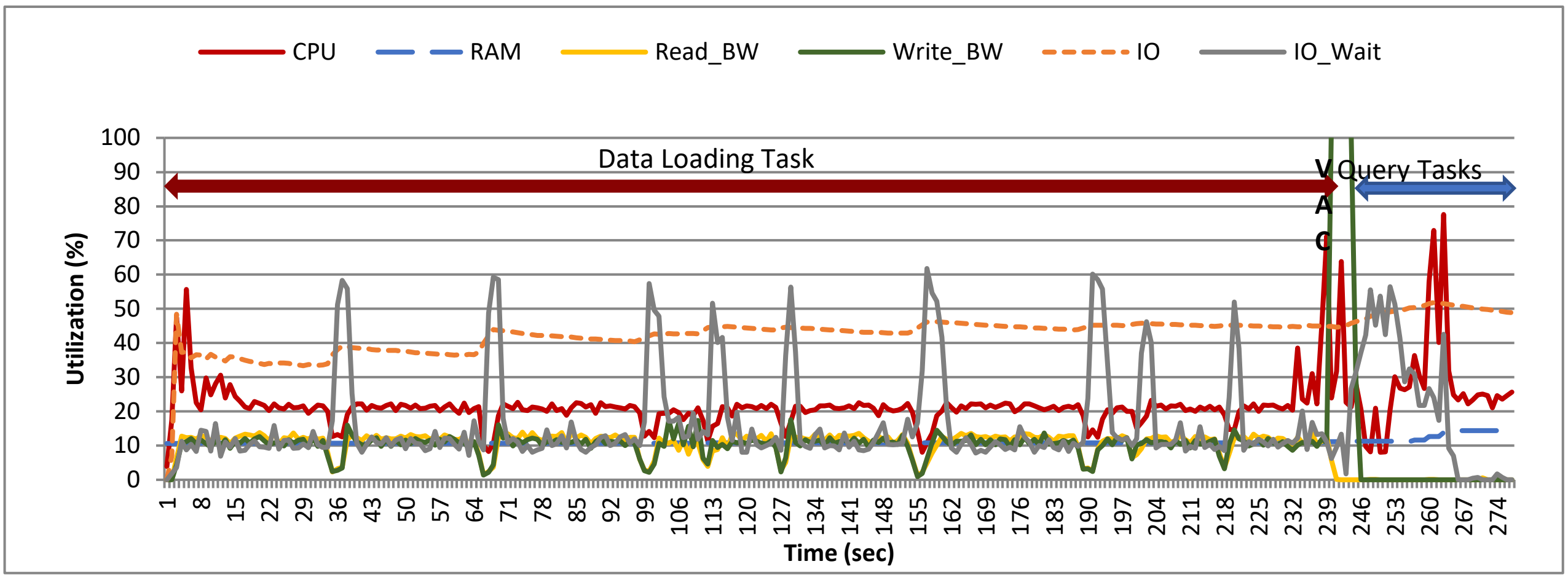

a) PostgreSQL

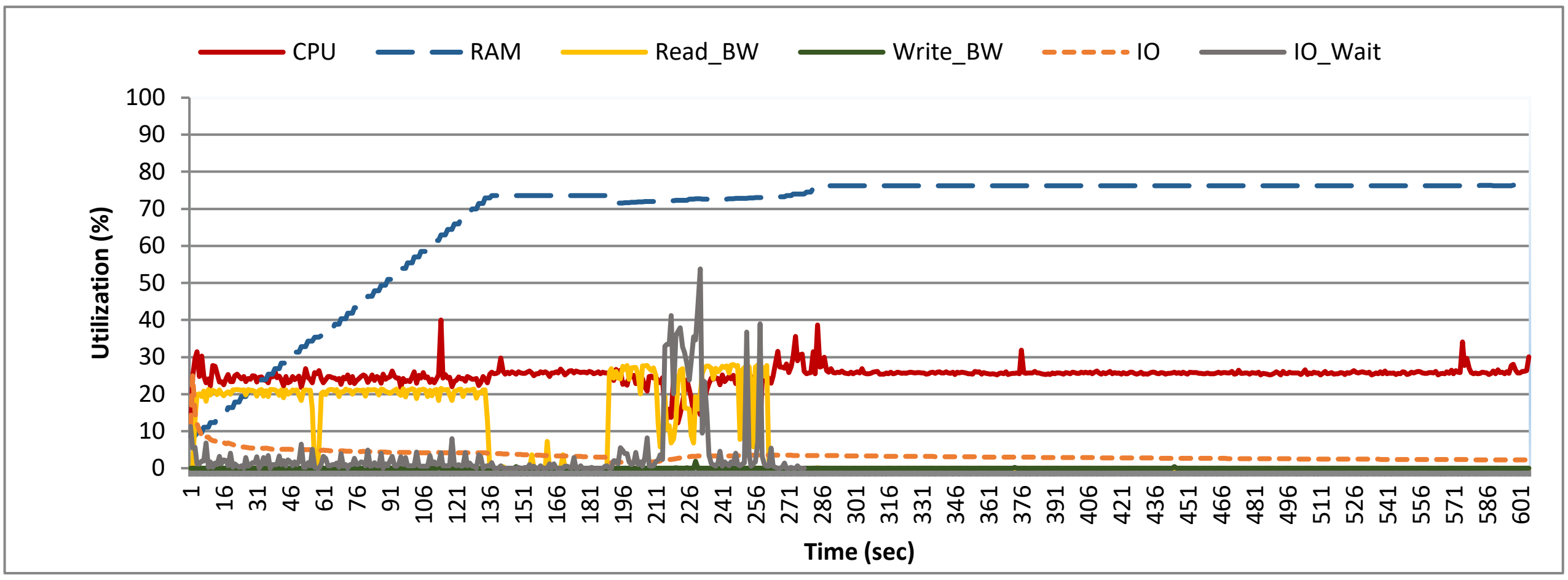

b) PostgresRAW

Fig. 7. Resource Utilization

Fig 7 b) shows resource utilization for the PgRAW tool. The tool accesses the data residing on the disk after the arrival of the query. For the first two queries up to 262 sec, PgRAW accessed the required data from disk, which increased IO utilization by more than 20%. The raw engine caches and indexes the data into the disk. Therefore, there are zero disk reads during the remaining WET. The raw engine does not write processed data back to disk, utilizing write bandwidth nearly 0%. However, the caching and indexing of raw data into main memory increased RAM utilization up to 76%. The CPU wait is almost zero once all data is cached in RAM. Therefore, query processing tasks utilized allocated CPU core completely, unlike PgSQL.

### 2) *Average Resource Utilization*

The average resource utilization of both tools is plotted in Fig 8. It can be observed that PgRAW utilized 3% more CPU than PgSQL due to 9x lower IO Wait. The RAM utilization of PgRAW is 6x more than PgSQL because of the caching and indexing of data in main memory. However, write IO utilization of PgRAW is zero due to no data loading, and read IO utilization is 2.5% less due to in-memory processing of queries. It can be noted that both tools required different resources. PgRAW utilized RAM most while IO resource is critical for PgSQL data processing. These results conclude that both tools require different resources to perform raw data query processing tasks. Therefore, once raw engine caches the data for query execution, the IO resource is available for data loading task.

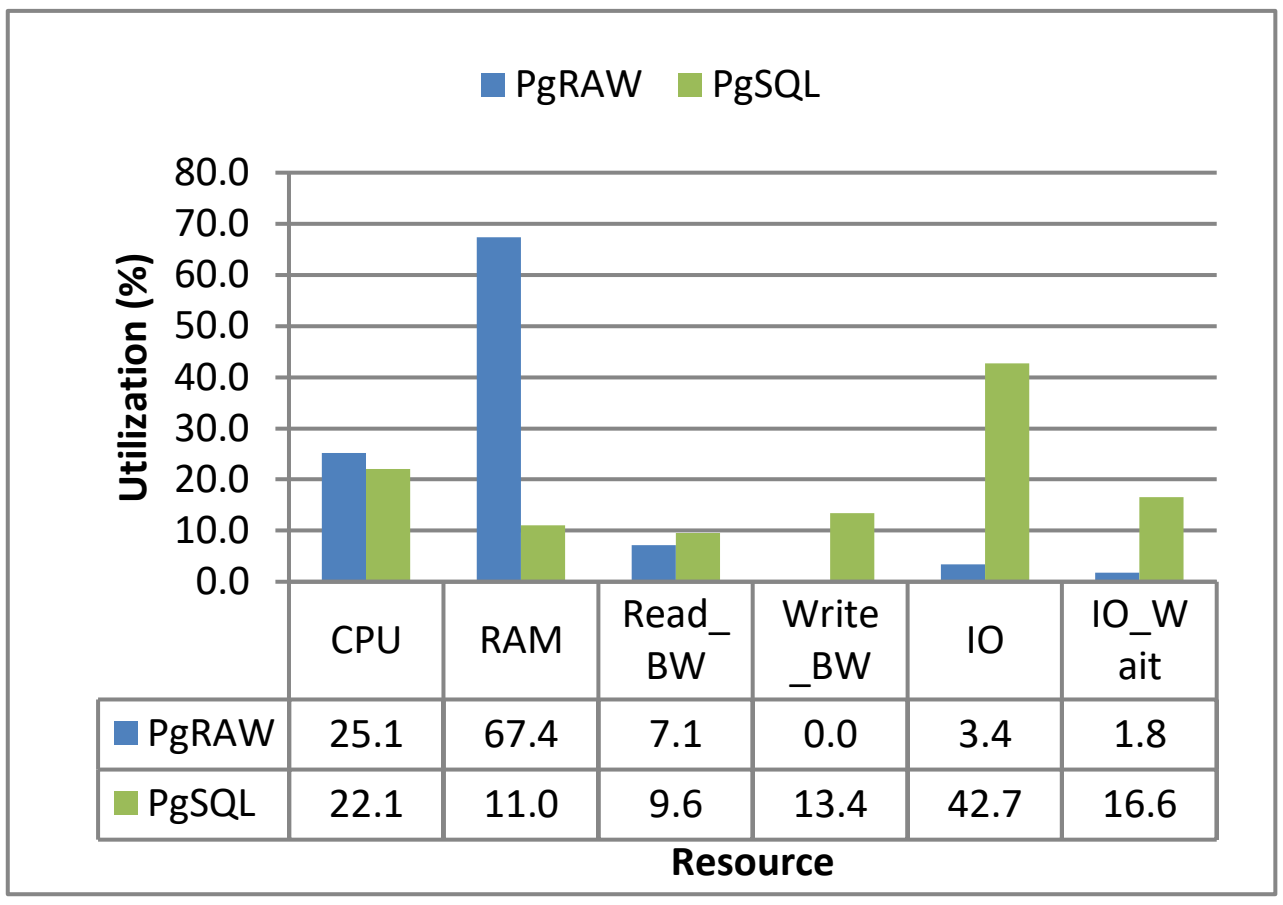

| | CPU | RAM | Read_BW | Write_BW | IO | IO_Wait |
|---|---|---|---|---|---|---|
| PgRAW | 25.1 | 67.4 | 7.1 | 0.0 | 3.4 | 1.8 |
| PgSQL | 22.1 | 11.0 | 9.6 | 13.4 | 42.7 | 16.6 |

Fig. 8. Average Resource Utilization

## E. *Data Scaling*

This section discusses the data scaling experiment results for both tools. Most researchers experiment with datasets smaller than the main memory size [19]. However, the experiments performed in this section used 18.8GB dataset size, which is 2.8GB larger than the RAM size.

### 1) *PostgreSQL*

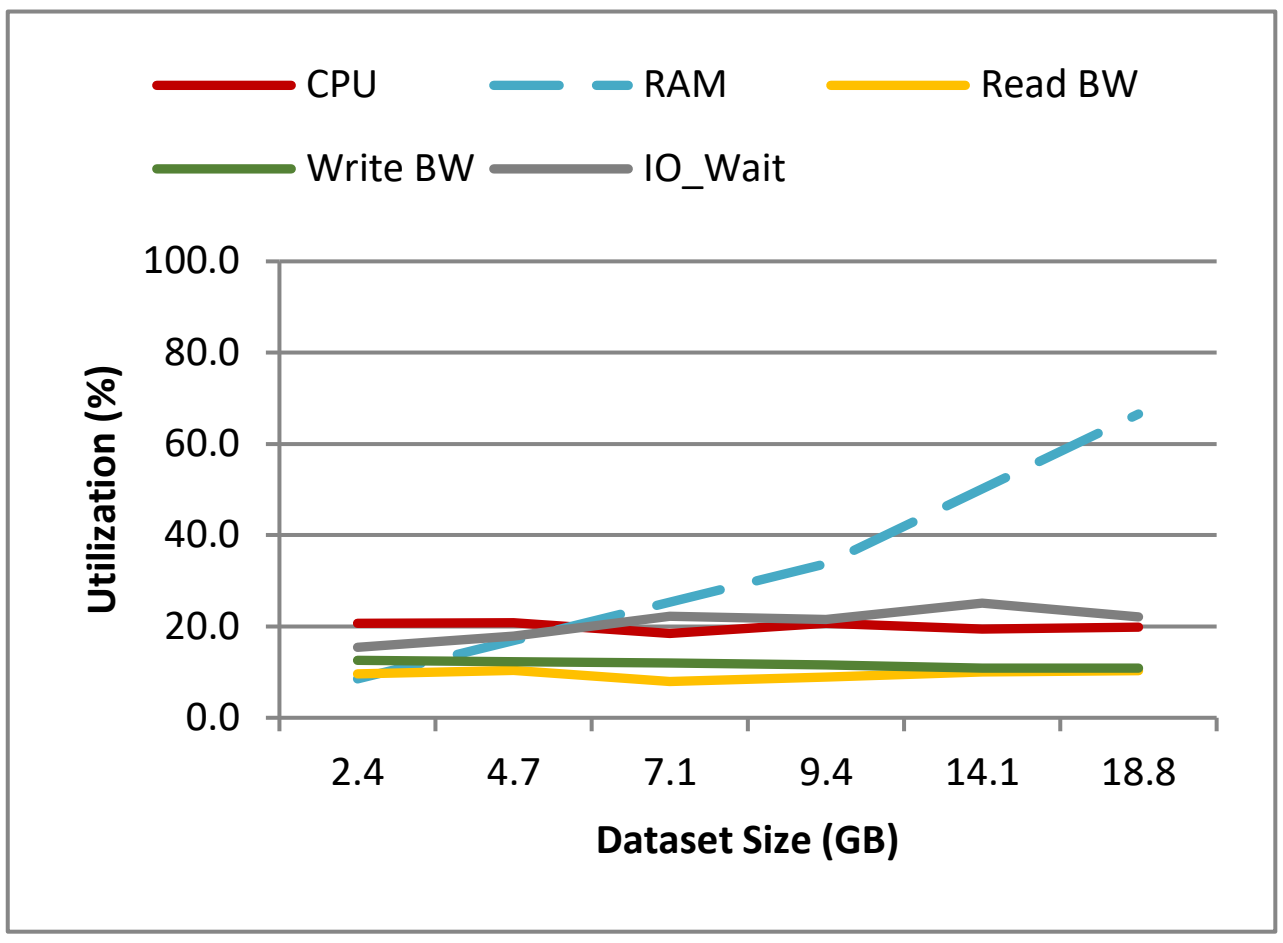

Fig. 9. Average Resource Utilization

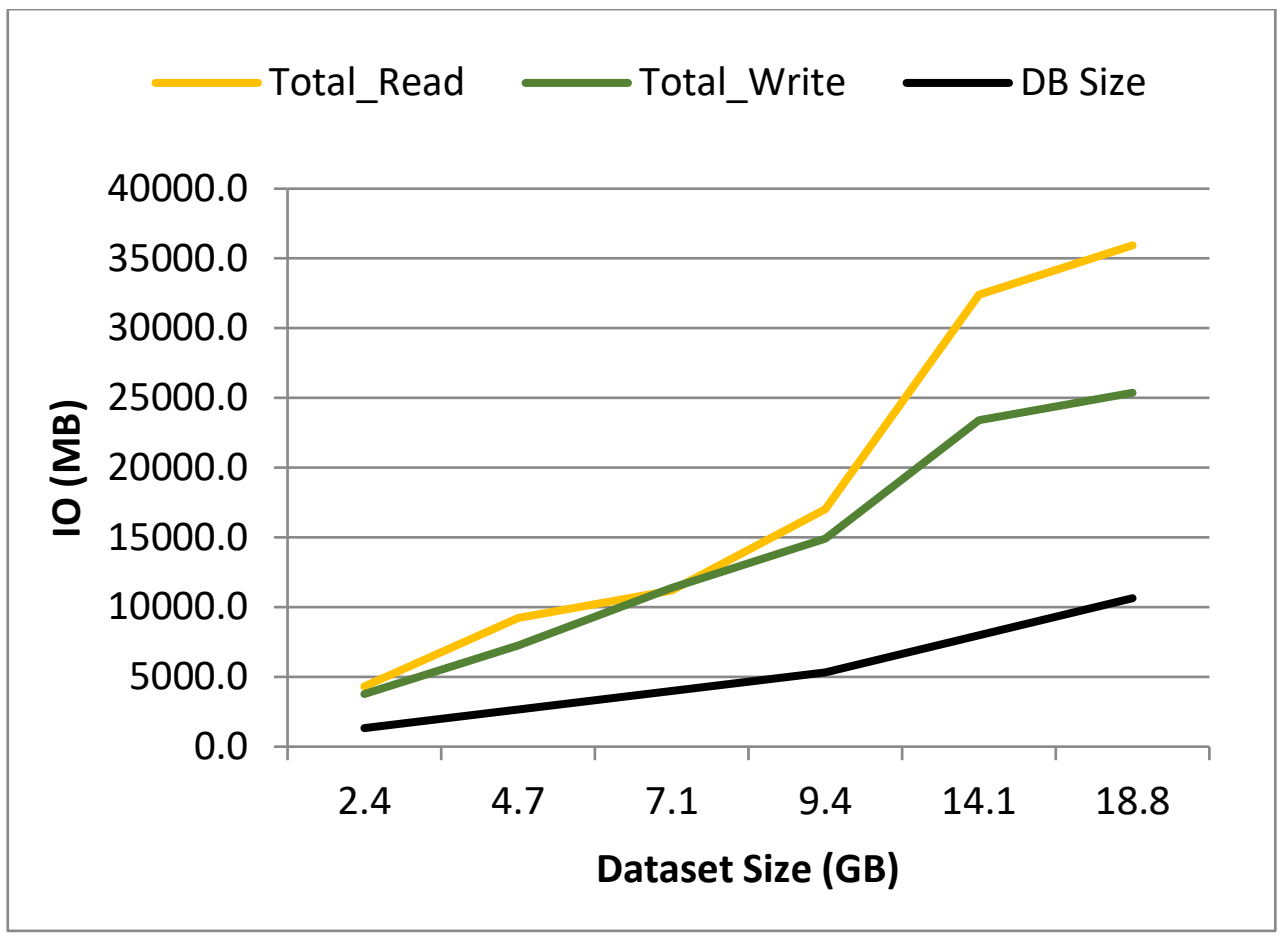

Fig. 10. Total IO Utilization

The data scaling experiments were performed by partitioning 18.8GB dataset file into eight smaller partitions, each containing 0.5M records. The 0.5M partitions were saved as CSV files, and these CSV files were loaded using COPY command followed by query execution tasks. Fig. 9 shows average resource utilization readings for the PgSQL processing SDSS dataset. It can be seen that CPU utilization is below 21% and decreases by 1-2% with the increase in dataset size. This happens because IO Wait increased from 15.4% to more than 21% for dataset sizes larger than 1M. The RAM utilization by *Postgres* process increased linearly from 0.4% to 1.6% for dataset sizes 2.4GB (0.5M) to 18.8GB (4M). The overall RAM utilization calculates the database files cached for processing queries. Therefore, the RAM utilization is plotted after including dataset size and Postgres process RAM utilization. It can be seen that RAM utilization increased from 8.5% to 66.6%, while IO Wait increased by 9.7% with dataset size. The CPU, Read & Write bandwidth utilization stayed below 21%. The *Postgres* process used only one core to load and query, following similar patterns shown in fig 7 a).

Fig 10 shows total IO utilization for PgSQL DBMS. The dataset size stored on a disk is 45% smaller than the raw file size. The total write is 1.5x, and the total read data size is 2.2x compared to the raw file size. The COPY command reads actual raw files having 4.7GB size and writes the compressed DB file of 2.6GB. The WAL writes logs of all inserted records in the non-compressed format before loading data, utilizing the write bandwidth by 1.5x. The Postgres process only reads the compressed database for query processing if not cached in RAM. The caching of database files in the main memory increases RAM utilization and Total Read size. The VACUUM, stat collection and check pointer processes use IO resources to arrange data on disk and collect statistical data.

### 2) *PostgresRAW*

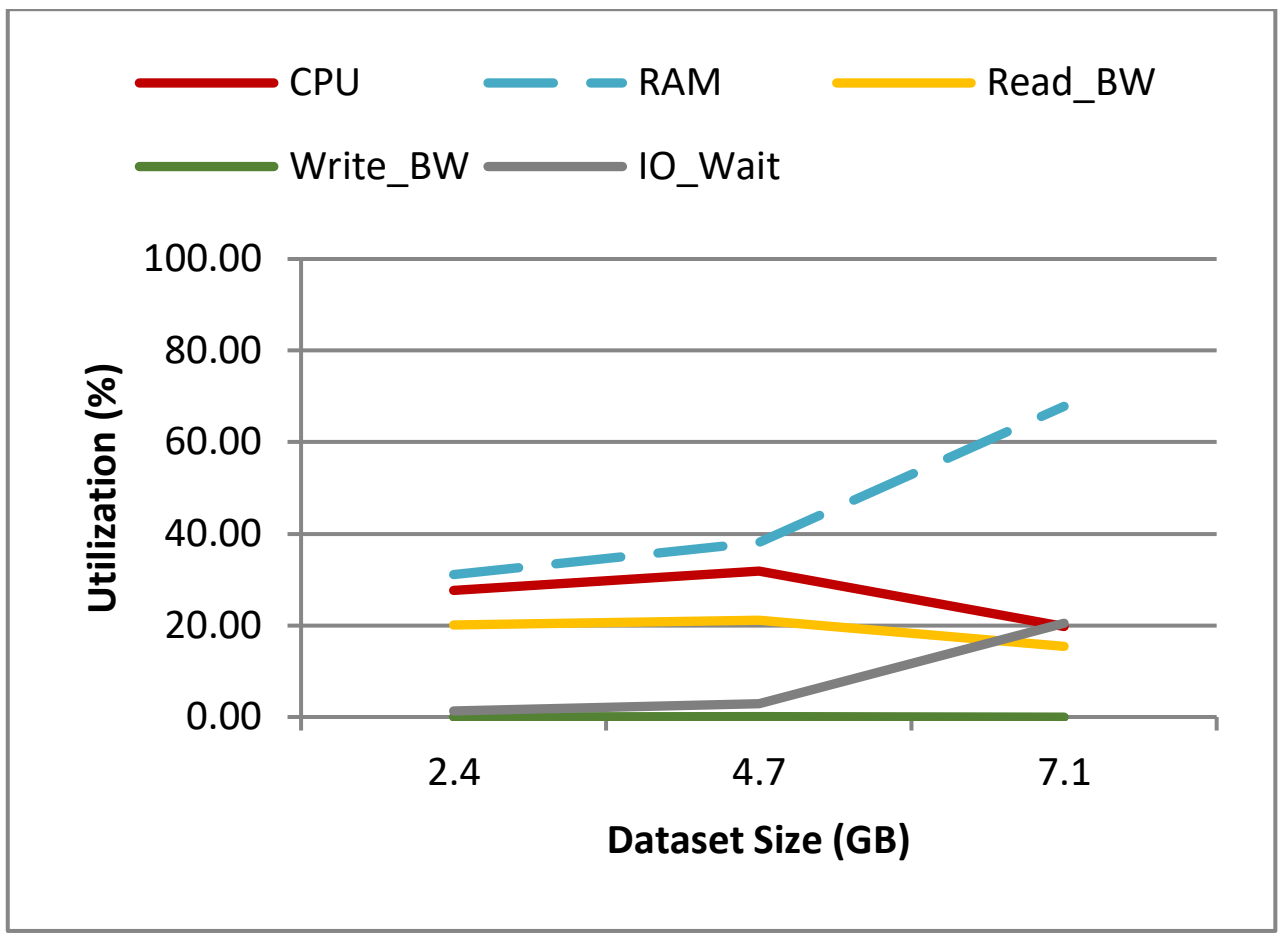

Fig. 11. Average Resource Utilization

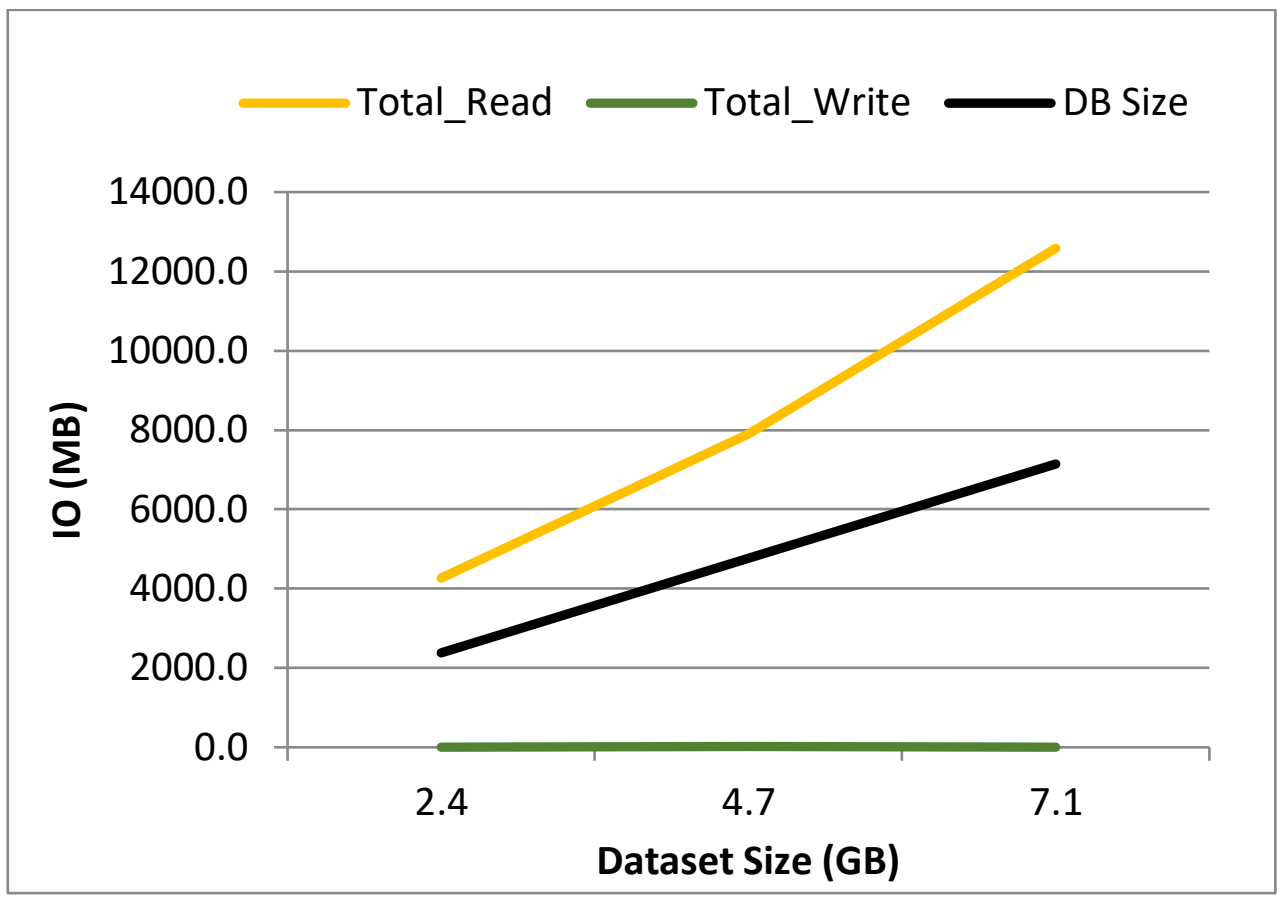

Fig. 12. Total IO Utilization

The experiments performed in this section use a single large raw data file to answer queries using in-situ processing PgRAW. Fig. 11 shows the resource utilization for data scaling experiments from 0.5M to 1.5M. The average RAM utilization increased up to 67% for 7.1GB (1.5M) file size. However, maximum RAM utilization had already reached 96.3% for 7.1GB (1.5M) & SWAP memory use had begun. The Postgres process had used 5-50% SWAP space in two different runs. The Postgres process went into sleeping mode, and other processes started crashing due to nearly 100% RAM utilization and high IO wait. Therefore, data scaling experiments had to be limited to 7.1GB. The high IO wait can be seen at 7.1GB point, which reduced CPU utilization by more than 12% compared to 4.7GB.

Fig. 12 shows total IO utilization by PgRAW tool to process SDSS dataset for 2.4GB (0.5M), 4.7GB (1M) & 7.1GB (1.5M) scaling experiments. It can be seen that the total read size is around 1.7x as compared to raw data file sizes. The detailed analysis of the *Postgres* process had shown that the total data read was exactly same as the raw file size. The additional data read might be due to OS and other processes caching the data from disk. The total data written to disk is less than 10MB which was written by RM threads of raw data query processing framework and stat collector process of Postgres.

The results analysis has shown that PgRAW required 64.4% around 10.3GB of RAM to run queries on 4.6GB of raw data, limiting the data scaling as it would have triggered swapping and increased WET. The RAM utilization of PgSQL is around 3x less, while PgRAW has almost zero write bandwidth requirements. The PgSQL is highly dependent on IO speed, while PgRAW is on RAM size. The CPU utilization of both tools is less than 31%. The results showed that PgSQL & PgRAW tools used only one CPU core for data processing tasks. It is because most traditional systems are designed with OLTP queries in mind, requiring sequential query task processing. Parallelizing data processing tasks for each query is difficult because it increases the overhead of identifying if a task is dividable without affecting the final result. The parallelization of a task needs division into smaller subtasks, distribution of subtasks, and combining intermediate results of those subtasks to produce the final result. The Cloudera and Hadoop based systems are highly parallelized systems. However, they required more resources to handle these additional parallelization tasks, increasing workload processing time.

The following section proposes the partitioning of raw data based on query classification and resource requirements.

### *F. Raw Data Partitioning*

The analysis of resource requirements of both tools established that PgRAW does not need IO resources after the data is cached and indexed in the main memory. While PgSQL needs IO resources longer to load data into DBMS. Both tools used a single CPU core. Therefore, CPU resources are available to handle both processes in parallel. However, high RAM utilization causes processes to stop in PgRAW, and high IO wait increases data loading time in PgSQL. The experiments performed in this section analyze individual query requirements to propose data partitioning to reduce WET.

#### *1) Query Resource Requirements*

Most database management systems cache the data for faster execution of queries. Fig. 13 shows the difference between Cold and Hot runs of simple queries executed using PgRAW of 1M records dataset. It can be seen that Q2, Q4, Q6, and Q10 require around ~134seconds to process raw data & provide results when processed data is not in main memory, similar to the data loading task of PgSQL. However, PgSQL required 109sec more to load data into DBMS before queries could be executed. The cold runs of Q7 provided the query results within 0.09sec, which is 2700x faster than PgSQL. It means 2700 queries like Q7 can be executed using PgRAW before PgSQL can finish data loading tasks. PgSQL caches the entire dataset into the main memory alongside data loading tasks. Comparison of Cold and Hot runs after for PgSQL had shown ~14sec difference. The difference between Cold & Hot runs of PgRAW is higher than PgSQL because PgSQL needs to cache the created database from the disk. On the other hand, when the PgRAW cache is cleared, it needs to process and index the raw data again.

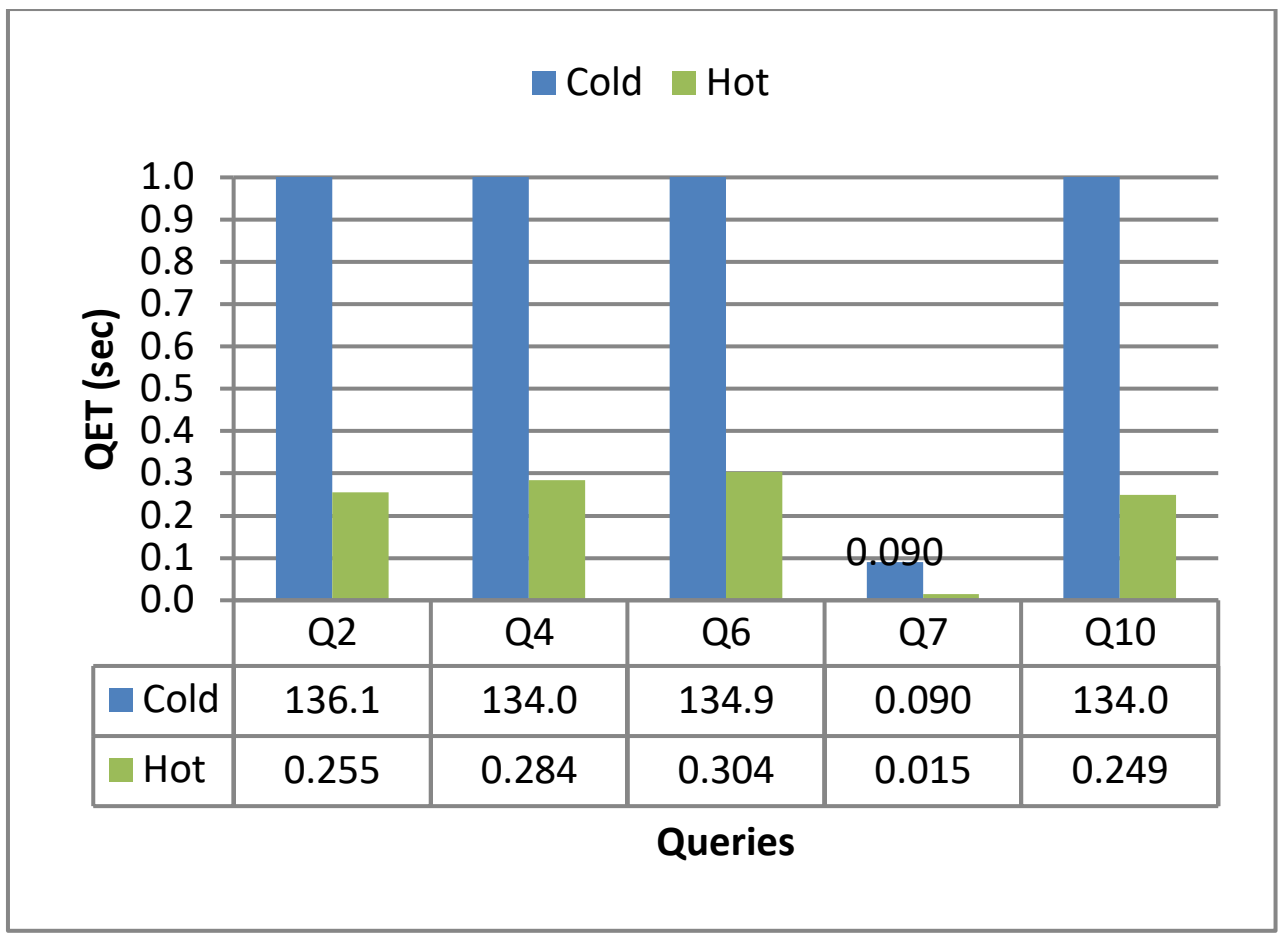

Fig. 13. Simple Queries QET: PgRAW

Fig. 14 displays RAM and IO resource utilization of all simple queries executed using PgRAW. It can be observed that Q7 requires less than 2MB of data to be read from disk, and RAM utilization was also less than 0.1%. The query Q7 is a sampling type query. Q7 needs to find top 10 records satisfying some conditions. Therefore, the query execution stops as soon as the Q7 result touches 10 records. At the same time, other simple queries required almost entire dataset to provide accurate results. Therefore, Q2, Q4, Q6, and Q10 utilized 10.3GB of RAM after accessing 4GB data from disk.

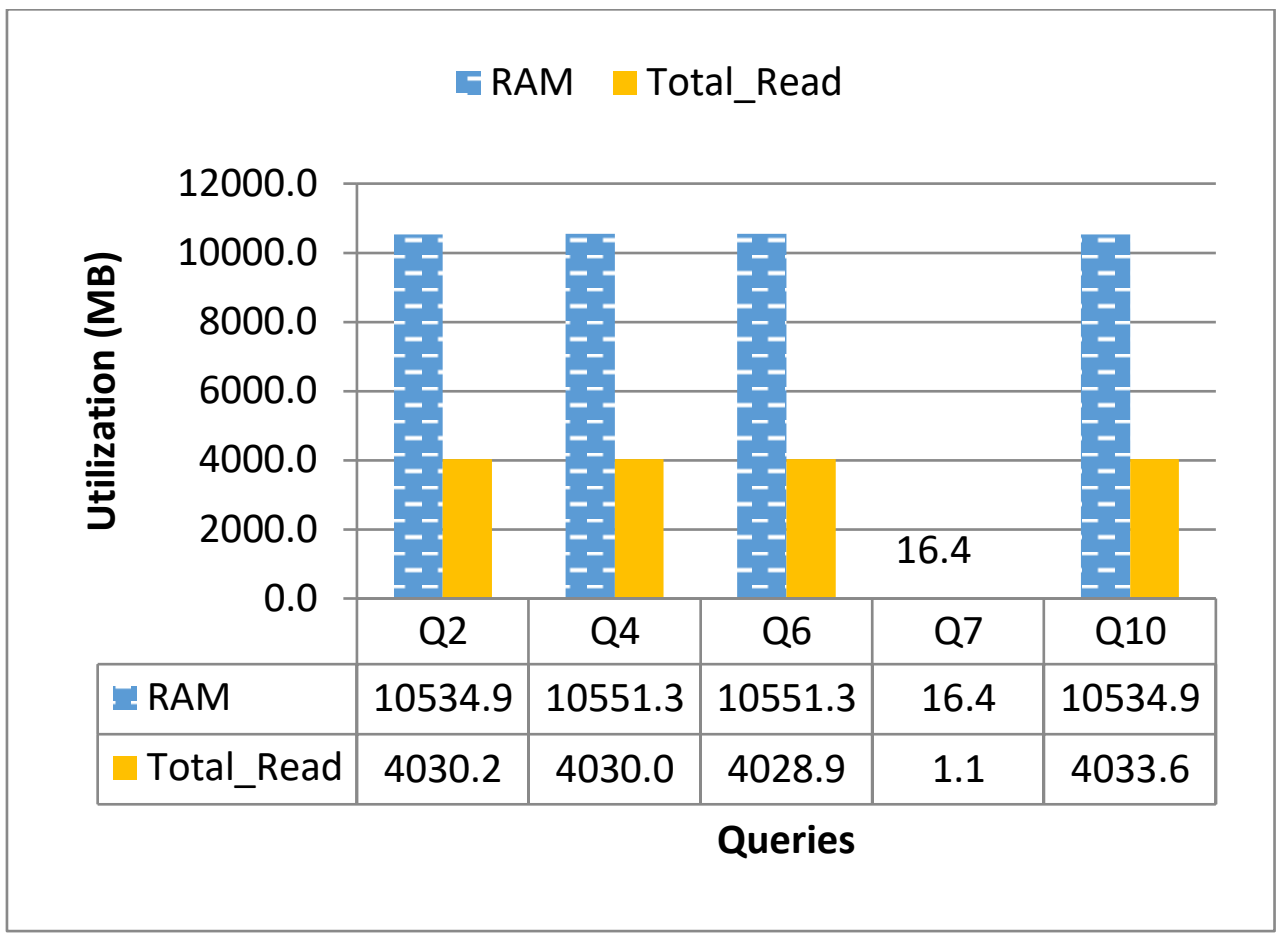

Fig. 14. Resource Utilization of Simple Queries: PgRAW

#### *2) Partitioning Techniques*

Researchers have proposed horizontal, vertical, and hybrid partitioning methods to partition the dataset based on query workload. This section tries to reduce resource utilization by identifying and processing only the required part of the dataset to answer queries for broad table datasets SDSS.

*a) Query Complexity Aware (QCA):* This partitioning technique proposes to partition the dataset based on simple and complex classification of workload queries. This partitioning technique is based on Fig. 6 results. The results have shown that PgRAW can answer simple queries faster than PgSQL. However, the first query is slow and takes a substantial time to process data existing in raw files. The QCA technique partitions the dataset by grouping attributes appearing in simple queries and keeping them in raw format. Fig. 15 shows a comparison of data to result time for simple queries. It can be seen that QCA[32] reduces WET by 93.78% compared to the original dataset and 53.02% reduction compared to workload-aware WA techniques [8], [18]. The workload-aware techniques load all workload attributes into a database if storage resources are available, while QCA keeps attributes of simple queries in raw format, which reduces the initial data processing time for PgRAW.

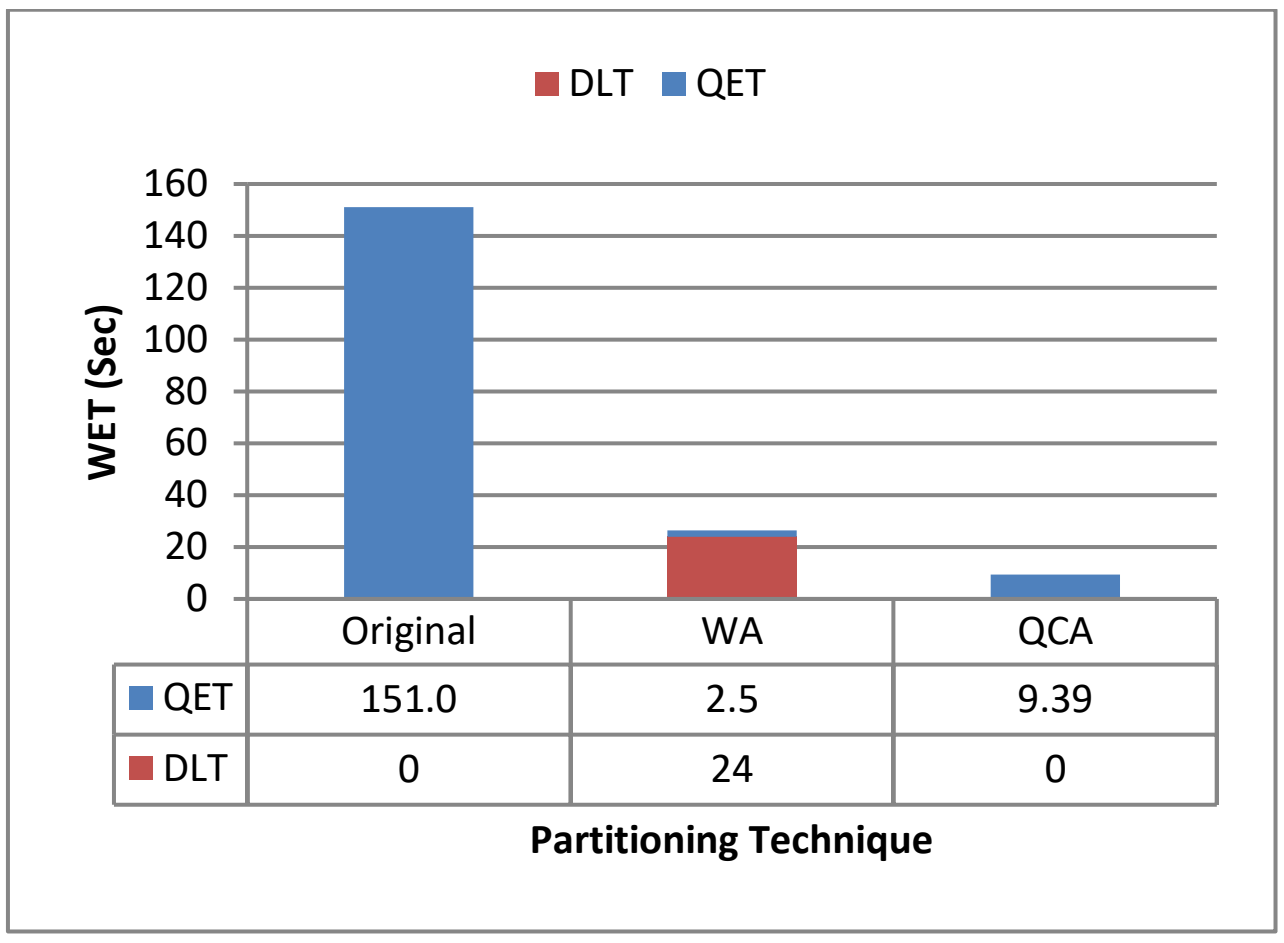

Fig. 15. WET: Simple Queries

*b) Resource Utilization Aware (RUA):*

This technique proposes partitioning the dataset based on the resource requirements of each query. Fig. 14 results show that PgRAW utilizes the minimum resource to answer sampling queries. The PgSQL needs to load the data into DBMS. Therefore, resource utilization of PgSQL is higher than PgRAW, as shown in Fig. 16. RUA technique partitions the raw dataset based on the attributes of sampling queries. The SDSS workload had only one sampling query, so the comparison shown in graph consists of single query QET-Q7.

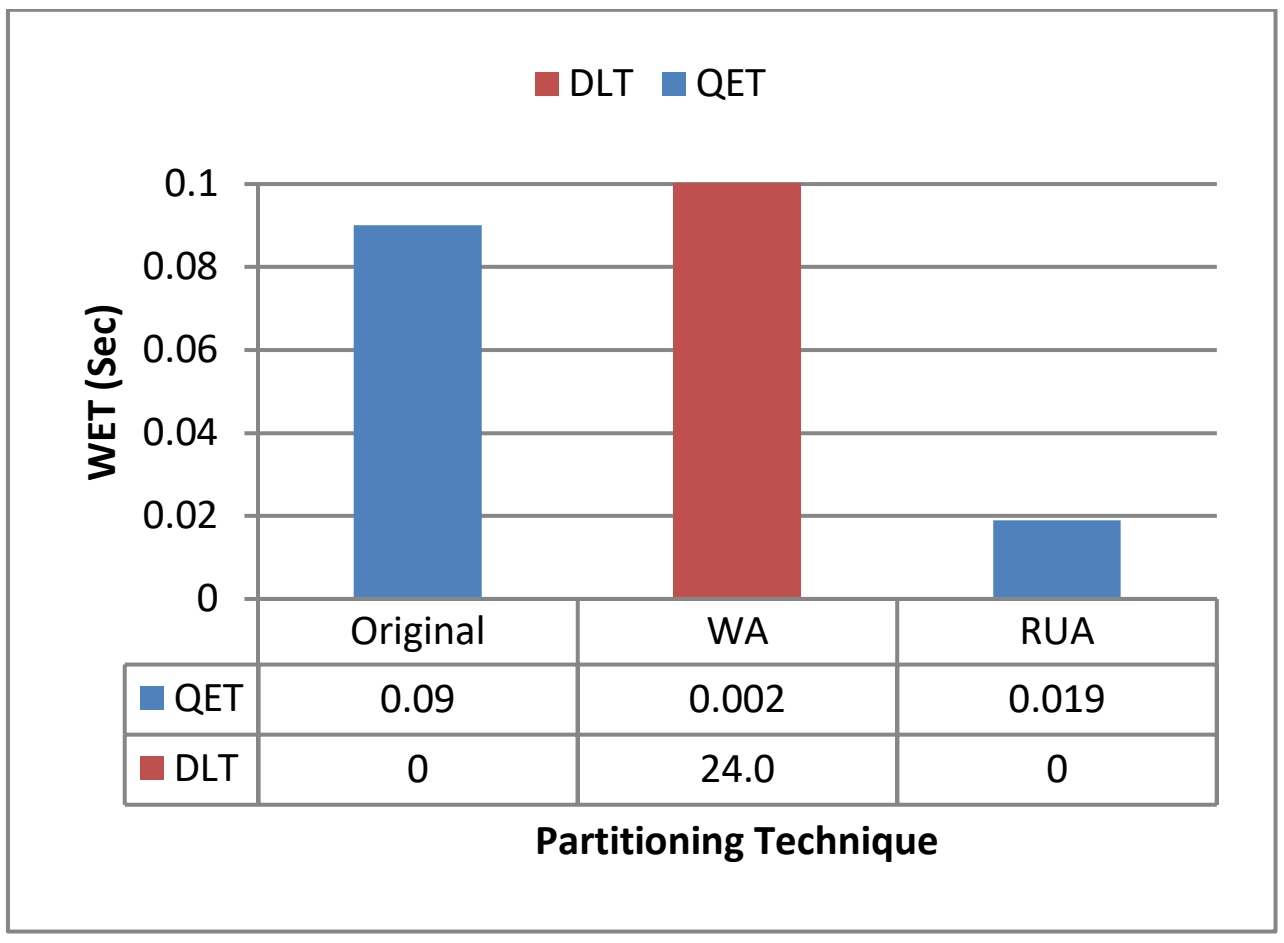

Fig. 16. WET: Sampling Query

*c) Comparison with state-of-the-art techniques:*

Table III. shows a comparison of state-of-the-art data processing and partitioning techniques for different parameters. The comparison results are generated using SDSS dataset and workload. W. Zhao[8], WSAC[18], QCA, and RUA use the vertical partitioning technique to partition the data, while PDC[19] uses the horizontal partitioning technique. The PDC proposed parallel processing of a single query on multiple nodes; however, it utilized a single CPU core out of 32 cores on a single node. Other techniques only proposed parallel loading of raw data to increase CPU utilization. PDC utilized 50% of RAM to build and cache data summaries of partitions. In comparison, WSAC proposes partitioning datasets based on available RAM or storage budget defined by database admin to cache database partition. The RUA considers the RAM utilization of each query for given DB tools. The QCA does not consider IO utilization to partition the dataset. PDC used global histograms to skip unwanted partitions. However, PDC needs to send data & receive results between nodes, increasing the internode communication. RUA proposes to compare all resource utilization, QET, and query type for given DB tools to partition the dataset among them to reduce overall resource utilization and data to result time. Zhao, WSAC, QCA & RUA techniques can eliminate internode communication by replicating required partitions. The percentage of data loaded, replicated, and accessed from raw has been calculated based on the number of attributes in partitions compared to the original dataset. It can be seen that QCA and RUA techniques reduced the amount of loaded data by 4.9% & 1.4% while replication by 10% & 8.8% compared to state-of-the-art workload aware raw data partitioning WA techniques W. Zhao[8] & WSAC[18].

## VII. Conclusion

The paper proposed Resource Monitoring framework (RMF) and its implementation using *top* and *iotop* tools. The CPU, RAM & IO resource monitoring modules of RMF are integrated into the raw data query processing framework to analyze resource utilization patterns of different tools and data processing tasks. The analysis of monitored resources indicated that PostgresRAW could efficiently process simple queries. However, PostgresRAW is not optimized to answer one or more JOIN complex queries. Therefore complex queries need to be answered using PostgreSQL. PostgresRAW and PostgreSQL utilize different resources, which can allow loading of raw data in parallel to raw data query processing. The detailed analysis of data scaling experiments has shown that PostgresRAW can only process datasets less than 50% of the RAM size. The paper proposed Query Complexity Aware (QCA) partitioning technique based on query classification and QET time to reduce WET. The QCA reduced WET of simple query workload by 93.78%. In comparison, the Resource Utilization Aware (RUA) technique uses individual query resource requirements to find resource-efficient ways to process raw data. RUA reduced QET of sampling type queries by 78.9% compared to the original dataset execution in PostgresRAW. It can be concluded that resource monitoring allows us to find different data processing patterns which can be used to partition and distribute data for efficient processing. Automated analysis of workload queries and resource utilization can be done in future work to implement QCA and RUA techniques. Explore possibilities of applying machine learning techniques to both techniques.

Table III. Comparison: Partitioning Techniques

| Technique | Partitioning | Multi-Formats | DB Data % | Raw Data % | Replication % | Inter-node Comm. | Multi-Node Data Loading | Overhead of Partial Result Integration |
|---|---|---|---|---|---|---|---|---|
| HTAP [33] | - | DB-1, DB-2 | 200% | 0% | 100% | No | Yes | No |
| W. Zhao [8] | VP | Raw, DB | 10.6% | 0% | 10.6% | No | Yes | No |
| WSAC [18] | VP | Raw, DB | 10.6% | 0% | 10.6% | No | Yes | No |
| PDC [19] | HP | DB | 100% | 0% | 0% | Yes | Yes | Yes |
| QCA [32] | VP | Raw, DB | 6.7% | 5.9% | 1.8% | No | No | No |
| RUA | VP | Raw, DB | 9.2% | 1.6% | 0.2% | No | No | No |